# Specific heat and magnetization of a ZrB$_{12}$ single crystal: characterization of a type II/1 superconductor.

Yuxing Wang[1], Rolf Lortz[1], Yuriy Paderno[2], Vladimir Filippov[2], Satoko Abe[1], Ulrich Tutsch[1], and Alain Junod[1]

[1]*Department of Condensed Matter Physics, University of Geneva, 24 quai Ernest-Ansermet, CH-1211 Geneva 4, Switzerland*
[2]*Institute for Problems of Material Science, National Academy of Sciences of Ukraine, 03680 Kiev, Ukraine*

PACS number: 74.25.Bt, 74.25.Ha, 74.25.Op

**Abstract**

We measured the specific heat, the magnetization, and the magnetoresistance of a single crystal of ZrB$_{12}$, which is superconducting below $T_c \cong 6$ K. The specific heat in zero field shows a BCS-type superconducting transition. The normal- to superconducting-state transition changes from first order (with a latent heat) to second order (without latent heat) with increasing magnetic field, indicating that the pure compound is a low-k, type-II/1 superconductor in the classification of Auer and Ullmaier [J. Auer and H. Ullmaier, Phys. Rev. B **7**, 136 (1973)]. This behavior is confirmed by magnetization measurements. The *H-T* phase diagram based on specific-heat and magnetization data yields $H_{c2}(0) = 550$ G for the bulk upper critical field, whereas the critical field defined by vanishing resistance is a surface critical field $H_{c3}(0) \sim 1000$ G.

## 1. Introduction

The discovery of superconductivity in $MgB_2$ at $T_c \cong 39$ K has revived interest in the abundant class of metal-boron compounds.[1] Among dodecaborides, $ZrB_{12}$ stands out by its relatively high critical temperature $T_c \cong 6$ K.[2] Early specific-heat studies evidenced a high Debye temperature $\theta_D(0) = 930$ K, together with the presence of low energy modes in the phonon spectrum.[2] This favorable situation did not seem to be sufficient to account for the relatively high $T_c$ since the density of states at the Fermi level was found to be particularly low.[2,3] Recently, owing to the availability of large, high-quality single crystals, new studies were initiated using in particular modern spectroscopies.[4-8] In spite of the quality of the samples, some results obtained by different investigators were found to be inconsistent. For example, the upper critical field $H_{c2}(0)$ was reported to be ~1500 G according to transport and radio-frequency susceptibility measurements,[8] 1120 G according to point-contact spectroscopy,[5] but only 390 G according to DC magnetization.[6] It was also debated whether $ZrB_{12}$ is a type-I or type-II superconductor.[6] Enhanced pairing at the surface was considered.[6,7]

This motivated us to carry out new specific-heat experiments on this compound, in particular as a function of magnetic field. We recall that such thermodynamic measurements provide reliable bulk information, much less susceptible to surface or filamentary superconductivity than magnetic susceptibility and resistivity measurements. In this article, we study the *H-T* phase diagram based on specific-heat data measured on high-quality single crystals, and compare it with magnetization and magnetoresistance data measured on the same samples. Our results reveal that $ZrB_{12}$ belongs to the category of marginal superconductors, whose Ginzburg-Landau-Maki parameter $k_1$ [9] is close to $2^{-\frac{1}{2}}$ and changes with temperature. Thus the type of superconductivity crosses over from type-I near $T_c$ to type-II/1 (in the classification of Auer and Ullmaier [11]) below $\sim T_c/2$. The vortex distribution in the type-II/1 regime is influenced by attractive interactions between flux lines, and may lead to an intermediate mixed state. We present for the first time a detailed characterization of such a superconductor using relaxation and AC specific heat techniques versus temperature and field.

## 2. Experimental details

Single crystalline rods of $ZrB_{12}$ were grown by the Kiev group using a high frequency induction zone melting process.[12] A large piece $4.7 \times 4.8 \times 2.9$ mm$^3$ (ZB1, 210.5 mg), shaped by spark cutting followed by etching in a boiling $HNO_3:H_2O$ 1:1 solution, was used for specific heat measurements by the relaxation technique. The <100> axis was normal to the largest face. A second sample with a low demagnetization factor along the <100> direction (ZB2, ~5.3 mm × 0.33 mm$^2$, 6.5 mg) was cut with a diamond saw. It was used for AC specific heat, magnetization, and resistivity measurements.

The specific heat of sample ZB1 was measured in the temperature range 1.2 – 8 K using a "long relaxation" technique.[13] In this method, the instantaneous specific heat is obtained every 40 ms during a transient heating or cooling period which lasts typically 1 to 60 s. The total temperature variation during this transient is 1 to 2 K.

For sample ZB2, AC calorimetry was used.[14] This technique provides a higher data density and a better resolution for small samples; furthermore the specific heat can be measured at constant temperature versus the magnetic field. However, an inherent limitation of this method is that a latent heat does not appear in full in the presence of hysteresis.[15] Supplementary heat-flow measurements at constant $dT/dt$ were found to perform better for this purpose, but are not reported here.[16] All experiments were performed in a cryostat fitted with a 14/16 T magnet. Its residual field was carefully zeroed before specific heat measurements by maximizing the superconducting critical temperature of a Pb wire. Other fields were then measured using a low-temperature Hall probe. The magnetization was measured in a Quantum Design MPMS-5 magnetometer. The resistivity of the sample was determined by the four-probe method and a DC current reversal technique; the current used, 6 mA, did not give rise to any overheating.

## 3. Results and discussion

*3.1 Specific heat in zero field and normal-state data*

The specific heat $C/T$ of sample ZB1 is shown in Fig. 1 as a function of temperature. The superconducting transition in zero field manifests itself by a sharp, BCS-like jump with a midpoint at 5.91 K. The bulk transition width does not exceed 30 mK; this upper limit is given by the width of the temperature intervals near $T_c$. At 0.5 T, the sample is in the normal state. By a short extrapolation of the normal-state $C/T$ curve from 1.3 K to $T = 0$, we obtain the Sommerfeld constant $\gamma_n = 0.34$ mJ/K$^2$gat. This value is somewhat lower than the earlier determination 0.44 mJ/K$^2$gat obtained by a zero-field extrapolation from $T_c$ to zero.[2]

The specific heat in the normal and superconducting state is used to evaluate the thermodynamic critical field $H_c(T)$, which is a measure of the condensation energy:

$$-\frac{1}{2}\mu_0 V H_c^2(T) = \Delta F(T) = \Delta U(T) - T\Delta S(T),$$

$$\Delta U(T) = \int_T^{T_c} [C_s(T') - C_n(T')]dT',$$

$$\Delta S(T) = \int_T^{T_c} \frac{C_s(T') - C_n(T')}{T'} dT'.$$

Here $F$ is the free energy, $U$ the internal energy, $S$ the entropy, and the indices $n$ and $s$ refer to the normal and superconducting state, respectively. The specific heat $C$ is given per gram-atom (17.0 g) and $V$ is the volume of one gram-atom (4.68 cm$^3$). The result is shown in Fig. 1. The thermodynamic critical field at $T = 0$ is $H_c(0) = 415 \pm 10$ G; the uncertainty essentially arises from the extrapolation from 1.3 K to $T = 0$. The shape of the curve deviates from the 2-fluid model in a way that is typical of weak-coupling superconductors.

*3.2 Specific heat versus magnetic field: type II/1 superconductivity*

Figures 2a and 2b show the specific heat of samples ZB1 and ZB2 in different magnetic fields. We recall that two different techniques were used, relaxation for ZB1 (Fig. 2a) and oscillating temperature for ZB2 (Fig. 2b). As already mentioned, the latter cannot show the full latent heat at $T_c(H)$ if the oscillation amplitude is smaller than the hysteresis, which explains why the specific heat rises to higher peak values in Fig. 2a compared to Fig. 2b. [15] An interesting feature of the "long relaxation" technique is that the specific heat can be measured successively during heating and during cooling through the superconducting transition. The initial state is field-cooled from above $T_c$. Some irreversibility is observed: the latent heat, i.e. the area below the specific-heat peaks, is larger in the heating curves than in the cooling curves; this is compensated by a slightly different specific heat at lower temperature. This phenomenon will be discussed below in terms of vortices and pinning. At first glance, the specific heat behaves as expected for a type-I superconductor. Generally speaking, the abrupt field expulsion that accompanies the transition from the normal to the Meissner state is responsible for the appearance of the latent heat, which must vanish at $T_c$ and $T = 0$; in the two-fluid approximation, a maximum is expected for $T/T_c = 3^{-1/2}$. In our data, this latent heat manifests itself as a specific-heat peak already visible for an applied field of 4 G; it only disappears when the field of the superconducting magnet is carefully zeroed. The width of the peak increases with the field. This is a consequence of the distribution of local fields and the characteristic lamellar structure which occur in the intermediate state.[17] Further detailed analysis of the curves measured on sample ZB1 is made difficult by demagnetization effects.

This problem is avoided by using sample ZB2, whose demagnetization factor is $D \approx 0.03$, ten times smaller than for ZB1. AC specific heat measurements (Fig. 2b) show great detail compared to relaxation calorimetry. Peaks are sharp in low fields, and the specific heat below the transition appears to be independent of the magnetic field up to about 150 G. This is a characteristics of type-I superconductivity, whereas vortex cores would give rise to a field-dependent specific heat in the superconducting state of a type-II superconductor. A crossover is observable in the vicinity of 4 K (corresponding to $H_c \approx 200$ G). At higher temperature (lower field), a peak develops at the superconducting transition, which is obviously of first order, whereas at lower temperature (larger field) the transition becomes continuous and takes the shape of a broadened step. The inset of Fig. 2b shows an expanded view of this crossover region. Based on the order of the thermodynamic transition along the $H_{c2}(T)$ line, one is lead to conclude that $ZrB_{12}$ crosses over from type I to type II with decreasing temperature.

An analogous behavior has been reported in magnetization experiments performed on superconductors with a value of the Ginzburg-Landau parameter κ slightly below $2^{-1/2}$, e.g. tantalum doped with nitrogen.[11,18-21] The Maki parameter $\kappa_1(T) \equiv 2^{-1/2} H_{c2}(T)/H_c(T)$, $\kappa_1(T_c) = \kappa$, generally increases with decreasing temperature, so that the initially type-I superconductor in the Meissner state acquires type-II character at lower temperature, i.e. flux lines enter the sample. An additional phenomenon arises in the latter situation: the attractive component of the interaction between vortices, which is negligible in superconductors characterized by small vortex cores on the scale of the penetration depth, starts to play a role. Upon increasing the field through $H_{c1}(T)$, the Shubnikov phase forms with an equilibrium lattice parameter $d_0 < (\sqrt{3}\Phi_0/2\mu_0 H_{c1})^{1/2}$ determined by the interaction potential rather than the external field. This leads to the persistence of a magnetization jump $B_0$ on the $H_{c1}(T)$ line. This marginal situation was named type II/1 superconductivity.[11, 19-21] If the Ginzburg-Landau parameter κ is increased (e.g. by doping), a critical value κ$_{cr}$ is reached where the induction jump $B_0$ vanishes, the transition at $H_{c1}(T)$ becomes of higher order, and the net flux line interaction is of the usual repulsive character. Beyond κ$_{cr}$ the usual type-II/2 region is entered. The marginal case of type II/1 superconductivity has been observed for κ$_{cr}$ up to about 1.1.[19]

Type II/1 is distinct both from type- I, for which the induction jump is given by $B_0(T) = \mu_0 H_c(T)$, and type-II/2, for which $B_0 = 0$. In the type II/1 region, the discontinuous increase of the flux density $B_0 > 0$ is responsible for the presence of a latent heat. Therefore in $H > 0$ the transition of type-I superconductors is of first order at $H_c(T)$, that of type-II/1 superconductors is of first order at $H_{c1}(T)$ and of second order at $H_{c2}(T)$, whereas type-II/2 superconductors only have second-order transitions.

If the demagnetization factor $D$ of a type II/1 superconductor is finite, an *intermediate mixed state* where Meissner regions coexist with Shubnikov regions can be observed. This was proven by a number of experiments either from magnetization measurements [22] or from neutron diffraction or decoration results. [11, 19, 21-25]. This coexistence extends over the field interval $(1-D)H_{c1} < H < H_1 + DB_0/\mu_0$.[24] In this two-phase domain, Shubnikov regions characterized by a *constant* vortex spacing $d_0$ gradually occupy a larger part of the sample cross-section as the external field rises. At lower fields, the superconductor is in the Meissner phase and behaves as a type-I superconductor. At higher fields, the superconductor is in the Shubnikov phase and type-II behavior prevails.

The data shown in Fig. 2a and 2b can be understood within this framework. The crossover from type-I to the type-II regime occurs near 4.0 to 5 K, i.e. the transition from the homogeneous Meissner state to the normal state only occurs for fields below 180 to 200 Gauss. At lower temperature / higher field, ZrB$_{12}$ enters the type-II/1 regime. The downward specific-heat jump locates the upper critical field $H_{c2}(T)$ whereas the peak marks the first-order transition at $H_{c1}(T)$. Due to the existence of an intermediate mixed state, the latter is broader for the nearly cubic sample ZB1 (Fig. 2a) than for the thin rod ZB2 (Fig. 2b). The separation between the step and peak features is made clear in the example shown in Fig. 3: for $H$ = 244 Gauss, the midpoint of the specific heat jump determines $T_{c2}(H) \cong 4.1$ K, whereas the small specific heat peak determines $T_{c1}(H) \cong 3.97$ K. At higher field/lower temperature, the peak is smeared and only an upward step in the specific heat marks the location of $H_{c1}(T)$.

Figure 3 also evidences the history dependence of the specific heat. While for $H$ = 50 G, the field-cooling (FC) and zero field-cooling (ZFC) measurements are identical, implying ideal type-I behavior, on the contrary differences are observed in the type-II/1 regime at 244 G. Below 3.7 K, the ZFC specific heat in 244 G coincides with that in 50 Gauss, showing that the ground state is a Meissner state, in agreement with virgin magnetization curves (shown below in Fig. 5). However, the FC specific heat is higher by ~0.09 mJ/K$^2$gat at low temperature, indicating the presence of vortices with a superficial density $n > 0$. In the FC mode, the superconducting phase is entered from the tail of the magnetization curve, i.e. in the type II regime; pinning then freezes part of the vortices below $H_{c1}$. The vortex-core contribution to the specific heat is of the order of $C/T \approx n 2\pi\xi^2 \gamma_n$.[27] Therefore the observed shift, $\sim 0.26\gamma_n$ at $\sim 0.44 H_{c2}$, shows that the density of pinned vortices represents ~58% of that of the ideal Shubnikov phase in the same field. FC cooling in 500 G or in 244 G results in identical specific-heat curves (measured with increasing temperature in 244 G), suggesting that pinning centers are saturated. The origin of pinning may lie in the surface where EDX measurements have shown impurities.[28] In the metastable FC state, Meissner regions (giving rise to the specific heat peak) and Shubnikov regions (causing the $C/T$ shift at low temperature) coexist, similar to the intermediate mixed state observed in samples with a large demagnetization factor.

Distinct transitions at $H_{c1}$ and $H_{c2}$, and magnetic hysteresis appear more clearly in AC specific-heat curves measured during field sweeps at constant temperature (Fig. 4). The lattice contribution is constant during such runs. After subtracting it, all curves merge into the normal-state value $C/T = \gamma_n$ at high field. We first focus on the ZFC sweeps with increasing field. The specific heat is constant in the low field limit, as expected in the Meissner state. At higher field, the transition to the normal state manifests itself by both a peak and a step. These two features occur at the same field $H_c(T)$ near $T_c$, then move apart at lower temperature. This marks the crossover into the type-II region where the $H_{c1}(T)$ and $H_{c2}(T)$ branches separate from the thermodynamic critical field $H_c(T)$. Note that the specific-heat jump at $H_{c2}(T)$ may be positive or negative in $C(H)$ runs, in contrast to $C(T)$ runs where it is always negative. The amplitude of the peak decreases at low temperature, only a shoulder remains on the low field side of the anomaly at 1.4 K. This follows the expected decrease of the latent heat at low temperature at least qualitatively, but additional effects such as the decreasing thermal energy available to overcome pinning potentials may play a role. Furthermore it should be remembered that the AC specific heat underestimates the latent heat if the hysteresis becomes large relative to the temperature oscillation. This phenomenon was qualitatively confirmed by varying the AC heating power in a test at one temperature.

Sweeps with decreasing field show two features consistent with previous observations. The slope in the low field limit is positive, which is attributed to the contribution of vortex cores $C/T \approx n(H) 2\pi\xi^2 \gamma_n$. Specific heat peaks are absent at low temperature. This behavior, analogous to that of $(\partial M/\partial H)_T$ (see below), is attributed to undercooling, or pinning which freezes vortices into the low-field region.

*3.3 Magnetization measurements*

Magnetization measurements provide an independent confirmation of the crossover from type I to type II/1 superconductivity. Figure 5 shows the magnetization of sample ZB2, selected for its low demagnetization factor, as a function of the magnetic field. At 5.5 and 5 K, the $M(H)$ curve exhibits textbook-like type-I behavior:

$4\pi M = -H$ for $H < H_c$, followed by a discontinuous jump to $M = 0$. At $T \leq 4.6$ K, the magnetization follows again $4\pi M = -H$ at low fields, and deviates slightly just before undergoing an abrupt change $4\pi M_0 \equiv B_0 < H$ at a field we define as $H_{c1}(T)$. Beyond this point, the magnetization approaches $M = 0$ smoothly, more like a type-II superconductor, and vanishes (on the scale of Fig. 5) at a field we define as $H_{c2}(T)$. These features are similar to those published for prototype type II/1 superconductors, [10, 11, 18, 21] except for some rounding near the minimum of the magnetization.

A hysteresis curve measured in the type II/1 temperature region is shown in the inset of Fig. 5. The magnetization jump is absent in the decreasing field branch, consistent with the absence of a specific heat-peak in the FC conditions discussed previously.

*3.4 Phase diagram in the H-T plane*

Previous observations are summarized in the phase diagram (Fig. 6), showing:
- the thermodynamic critical field $H_c(T)$ obtained by integration of the difference between the specific heat in zero field and that in the normal state (green line);
- the position of the peaks (red circles) and onsets of the jumps (green diamonds) observed in "vertical" specific-heat runs at constant temperature;
- the position of the peaks (purple circles) and onsets of the jumps (blue diamonds) observed in "horizontal" specific-heat runs at constant field;
- the point at which the resistance vanishes (triangles), as discussed in Section 3.5.

Above $T^* \cong 4.7$ K, the various determinations merge with the thermodynamic critical field $H_c(T)$, thus establishing type-I superconductivity. Below $T^*$, first- and second-order transitions move apart on different branches, $H_{c1}(T)$ and $H_{c2}(T)$, respectively. The condensation energy in the vortex phase between $H_{c1}$ and $H_{c2}$ is much smaller than the total condensation energy (compare the areas in Fig. 5), so that $H_{c1}(T)$ must remain close to $H_c(T)$. This was also observed in Ta-N samples, see Fig. 7 of Ref. 19. In addition, superheating may slightly shift the apparent $H_{c1}(T)$ curve closer to $H_c(T)$.[21] The upper critical field is extrapolated to $H_{c2}(0) = 550$ G using a WHH fit.[29]

The upper critical field $H_{c2}(T)$ and the thermodynamic critical field $H_c(T)$ are related by $H_{c2}(T) = \sqrt{2}\kappa_1(T)H_c(T)$.[9] Broadening and background contributions make the determination of $H_{c2}(T)$ based on the position of the specific-heat step somewhat ambiguous; the most reliable determination of $H_{c2}(T)$ is obtained from the end of the magnetization tail at 2 K (Fig. 5). Based on this, one obtains $\kappa_1(2\text{ K}) \cong 0.87$. By definition of the crossover, $\kappa_1(T^*) = 2^{-1/2}$ at $T^* \cong 4.7$ K. Assuming a linear behavior as a first approximation, one finds $\kappa \equiv \kappa_1(T_c) \cong 0.65$ for the Ginzburg-Landau parameter.

*3.5 Surface superconductivity and third critical field*

The magnetoresistance, which was measured from 0 to 2000 G, is negligible in the normal state (Fig. 7). The superconducting transition remains well defined at all fields, in particular the zero resistance points which are plotted in the top curve of Fig. 6. This additional critical-field line obviously lies above $H_{c2}(T)$ as measured by specific heat or magnetization. Superconductivity is known to persist in clean samples in the region adjacent to an insulator/metal interface up to a field $H_{c3}(T) \cong 1.7 H_{c2}(T)$ parallel to the surface.[23] In the present case, the ratio between $H_{c3}$ and $H_{c2}$ is somewhat larger than 1.7, which suggests enhanced pairing at the surface. The volume involved in the surface transition is vanishingly small, so that no detectable specific-heat anomaly occurs upon crossing the $H_{c3}(T)$ line. However, by expanding the scale of the magnetization measurements shown in Fig. 5 by a factor of 500, one can evidence the onset of hysteresis at the point where resistance vanishes (Fig. 7, inset). The $H_{c3}(T)$ curve is almost perfectly linear and extrapolates (linearly) to $H_{c3}(0)$ = 1010 G, i.e. about $1.8 H_{c2}(0)$.

## 4. Conclusion

In this paper we concentrated on the magnetic configurations of ZrB$_{12}$ in the superconducting state. The detailed investigation of the specific heat, magnetization and magnetoresistance of single crystals reveals that clean ZrB$_{12}$ is a superconductor whose Ginzburg-Landau parameter $\kappa \cong 0.65$ is close to the border value $2^{-1/2}$, furthermore it significantly varies with temperature. As a consequence, this material crosses over from type-I to type-II/1 superconductivity at $T^*/T_c \cong 0.78$ as the temperature is lowered. It follows that the thermodynamic nature of the transition changes from first to second order when the field is increased. The phase diagram in the *H-T* plane was established, based on bulk determinations. Values of the upper critical field in excess of ~1000 G reported up to now in the literature are attributed to the third critical field $H_{c3}(0)$, which only affects the surface of the sample. Further studies of this interesting compound by optical and scanning tunneling spectroscopy are under way.[30]

**Acknowledgements**

This work was supported by the Swiss National Science Foundation through the national Centre of Competence in Research "Material with Novel Electronic Properties – MaNEP". The authors thank V. Gasparov and Ø. Fischer for fruitful discussions.

**Figure captions:**

Figure 1. Specific heat of sample ZB1 in the normal (open diamonds) and superconducting state (open circles). Insert: full squares, thermodynamic critical field $H_c(T)$ as obtained from the specific heat in the normal and superconducting state; line, a fit using BCS model.

Figure 2a. Specific heat of sample ZB1 ($D \sim 0.3$) measured by the "long relaxation" technique in different magnetic fields. From right to left: $H$ = 0, 4, 44, 76, 94, 144, 244, 344, 444, and 5000 Gauss. Red symbols: sweeps with increasing temperature. Blue symbols: sweeps with decreasing temperature

Figure 2b. Specific heat of sample ZB2 measured by AC calorimetry in different magnetic fields. From right to left: $B$ = 10, 20, 50, 70, 100, 120, 150, 170, 180, 190, 200, 210, 220, 230, 240, 250, 270, 300, 350, and 400 Gauss. Inset: expanded view from 170 to 400 Gauss. The arrows in the main frame point to the transition onsets used to define $H_{c2}(T)$, those in the inset point to the first upward deviation from the low-temperature behavior.

Fig. 3. FC and ZFC specific–heat measurements of sample ZB2 in 50 G (type I region) and 244 G (type II/1 region). An additional measurement at 244 G performed after field cooling in 500 G is shown in red.

Figure 4. AC specific heat of sample ZB2 at fixed temperature and variable field. Results for increasing field are shown by solid lines, results for decreasing field by dashed lines.

Figure 5. Virgin magnetization curves of sample ZB2 at fixed temperatures, from left to right $T$ = 5.5, 5.0, 4.6, 4.0, 3.0, and 2.0 K. The dashed lines are "guides to the eye" to show the flux jump in the Type-II/1 superconductor at $H_{c1}$. Inset: hysteresis loop measured at 4.6 K.

Figure 6. Phase diagram summarizing transition points obtained from specific heat and magnetoresistance. Red circles: specific heat peaks at $H_{c1}$ in field sweeps. Green diamonds: onset of specific heat steps at $H_{c2}$ in field sweeps. Purple circles: specific heat peaks at $H_{c1}$ in temperature sweeps. Blue diamonds: onset of specific heat steps at $H_{c2}$ in temperature sweeps. Triangles: the surface upper critical field $H_{c3}$ defined by $R = 0$ from the magnetoresistance measurements. $H_{c2}(0)$ is obtained from a WHH fit. The dashed lines are just guide to the eyes. Inset: $T/T_c$-$\kappa$ phase diagram showing the phase boundary between Type-I and Type-II/1 as well as between Type-II/1 and Type-II/2 superconductivity in the $T/T_c$-$\kappa$ plane,[11,19] the red arrow marks the position of the sample.

Figure 7. Resistance of sample ZB2 in various fields, from right to left 0, 49, 97, 146, 243, 340, 483, 631, and 2000 G. The field is parallel to the long axis of the sample and to the current flow. Inset: expanded view of the magnetization of the hysteresis loop measured at 4.6 K in the vicinity of $H_{c3} \cong$ 230 G.

Fig. 1

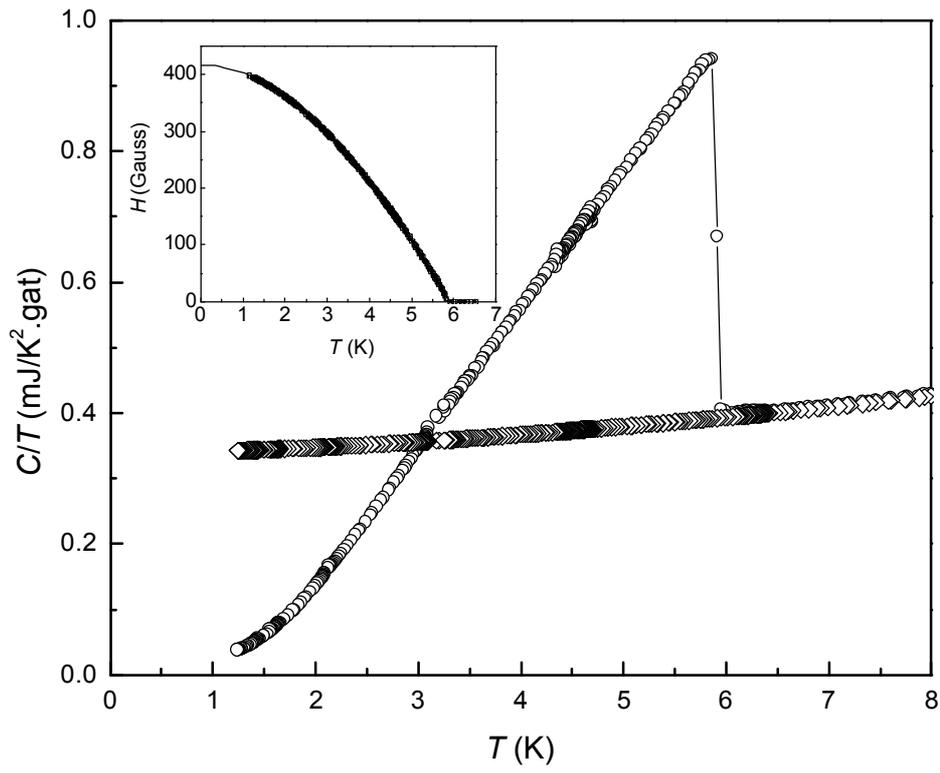

Fig. 2a

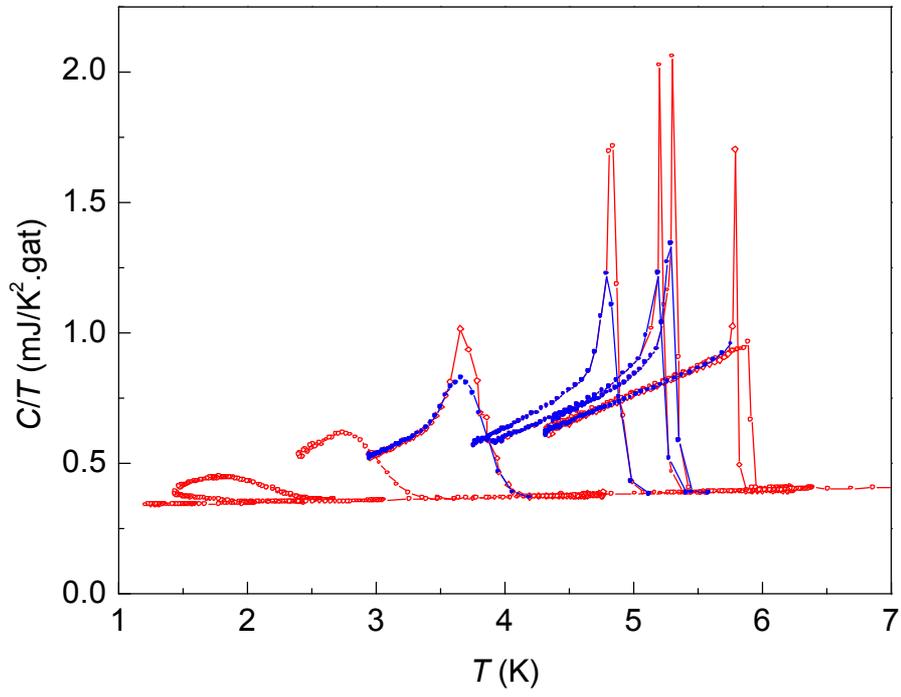

Fig. 2b

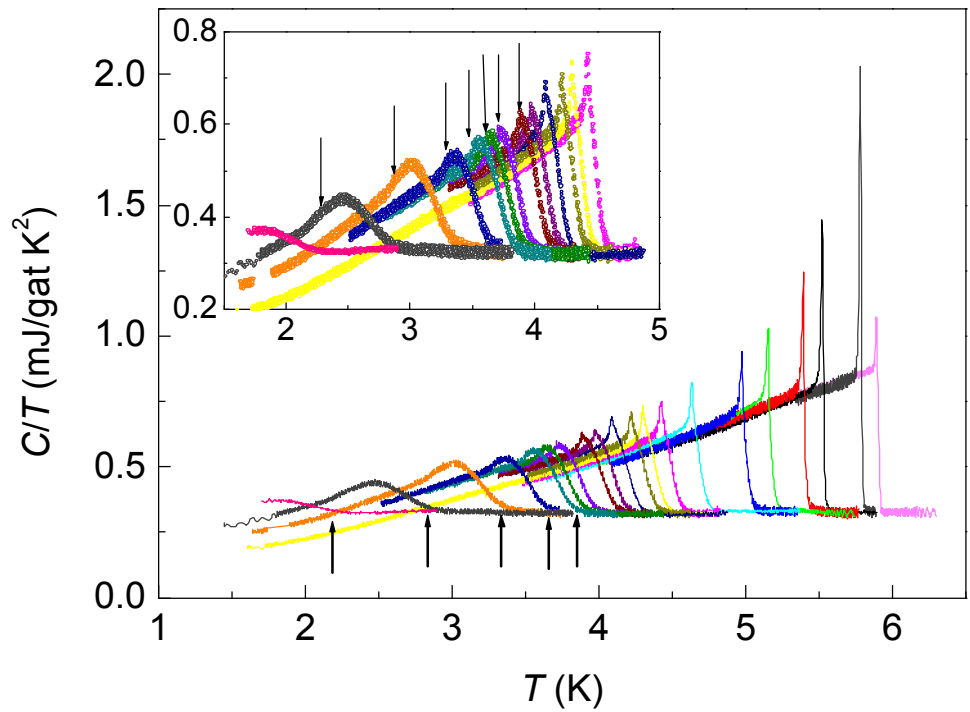

Fig. 3

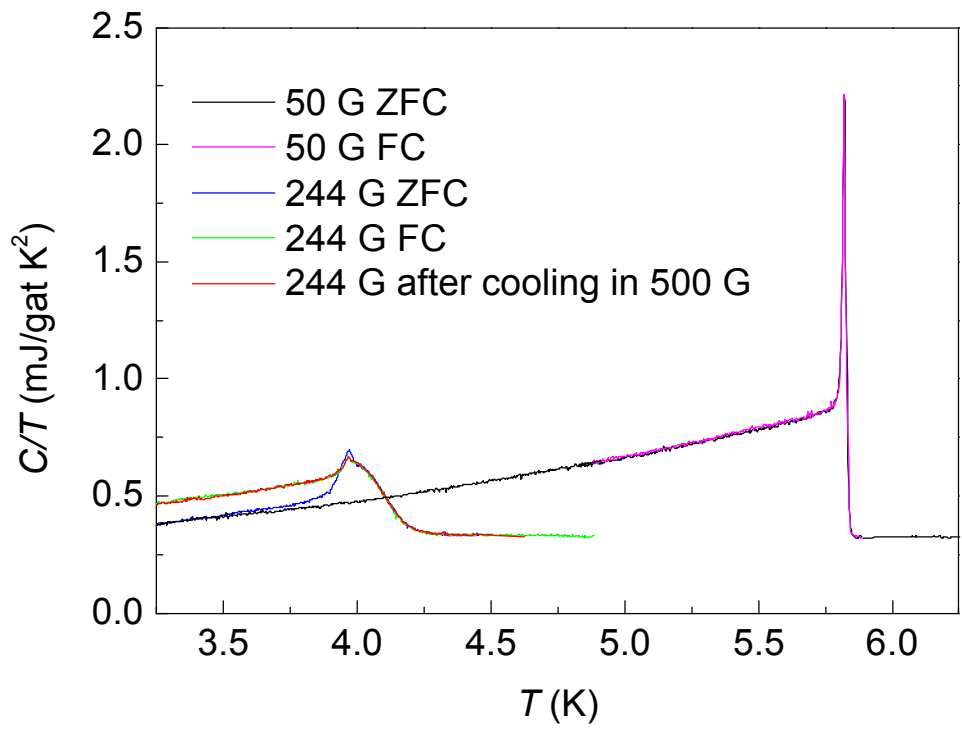

Fig. 4

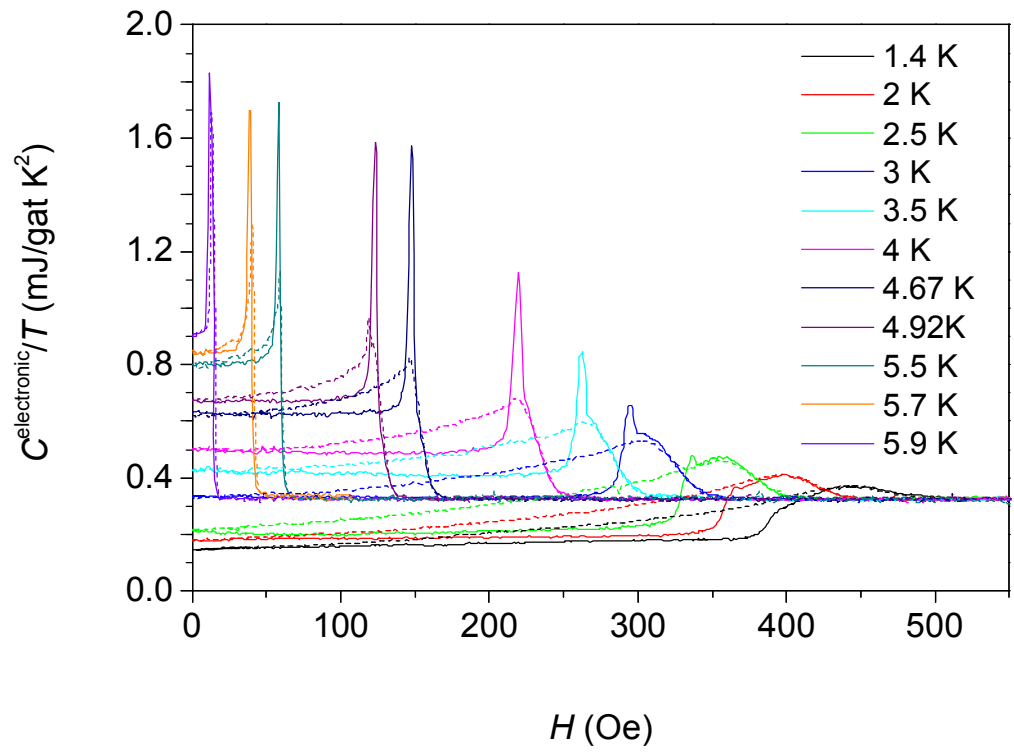

Fig. 5

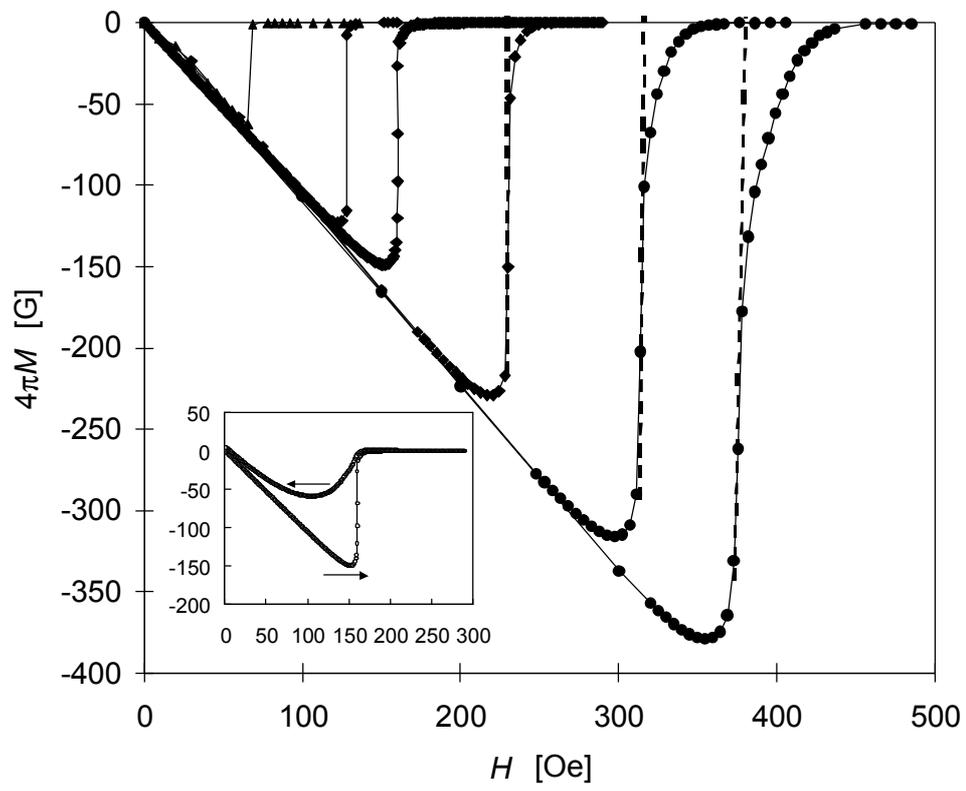

Fig. 6

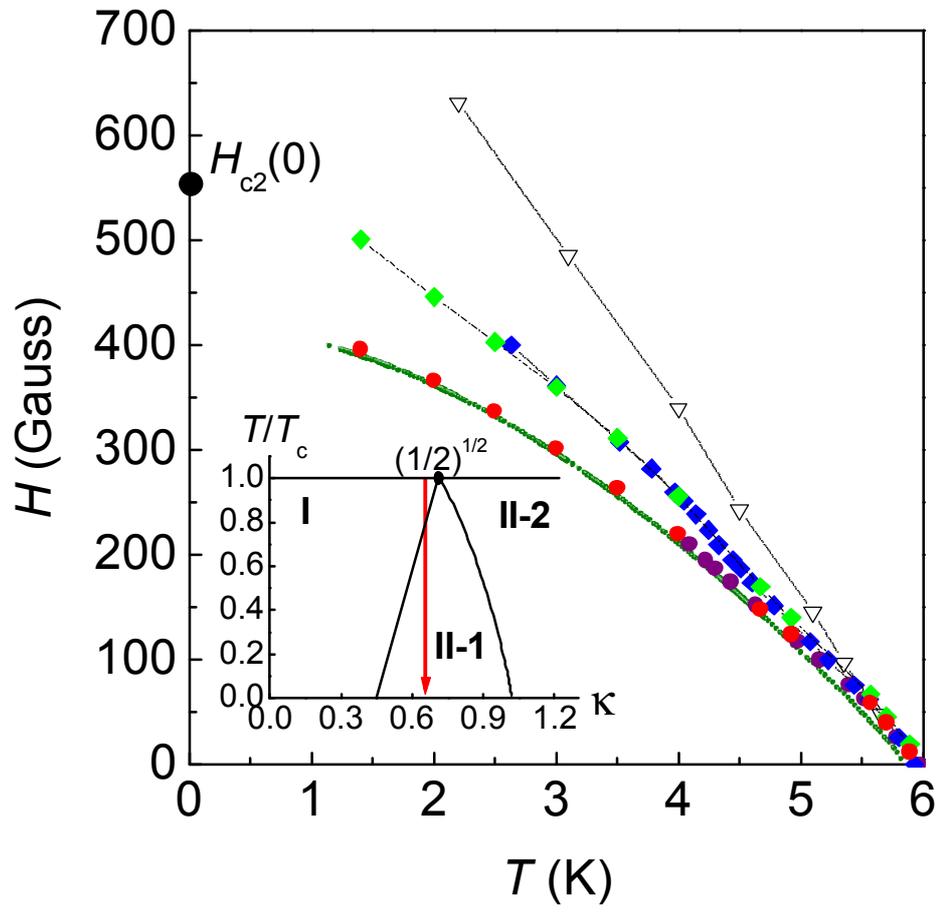

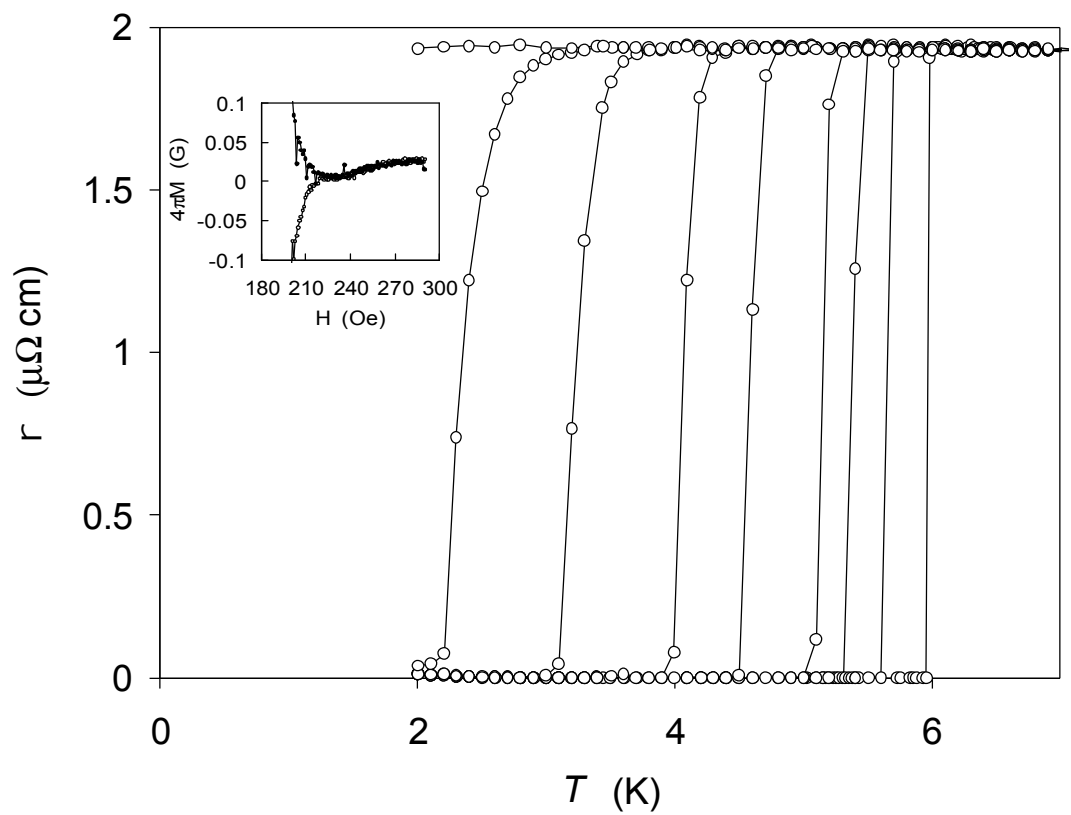
17